# Why determinism in physics has no implications for free will

Michael Esfeld

University of Lausanne, Department of Philosophy

Michael-Andreas.Esfeld@unil.ch

**Abstract**

This paper argues for the following three theses: (1) There is a clear reason to prefer physical theories with deterministic dynamical equations: such theories are maximally rich in information and usually also maximally simple. (2) There is a clear way how to introduce probabilities in a deterministic physical theory, namely as answer to the question of what evolution of a specific system we can reasonably expect under ignorance of its exact initial conditions. This procedure works in the same manner for both classical and quantum physics. (3) There is no cogent reason to take the parameters that enter into the (deterministic) dynamical equations of physics to refer to properties of the physical systems. Granting an ontological status to parameters such as mass, charge, wave functions and the like does not lead to a gain in explanation, but only to artificial problems. Against this background, I argue that there is no conflict between determinism in physics and free will (on whatever conception of free will), and, in general, point out the limits of science when it comes to the central metaphysical issues.

*Keywords*: classical mechanics, determinism, free will, functional reduction, Humeanism, physical laws, quantum mechanics

*1.    Determinism and probabilities in physics*

The central claim of this paper is that laws in physics, even deterministic laws, do not pose a threat to human free will. That claim is intended to come out as a consequence of considering the role of laws and probabilities in physics as well as an argument to the effect that a certain version of Humeanism, dubbed Super-Humeanism, offers the best metaphysical account of these laws as they figure in our physical theories. Therefore, the paper first goes into determinism and probabilities (this section), then considers the ontological status of the magnitudes that enter into the laws of physics (section 2) and finally draws conclusions for free will (section 3).

Atomism is the paradigm on which the success of modern science is based. It is the idea that matter is composed of tiny, indivisible particles. In fact, atomism is as old as philosophy, going back to the Presocratics Leucippus and Democritus. The latter is reported as maintaining that

> ... substances infinite in number and indestructible, and moreover without action or affection, travel scattered about in the void. When they encounter each other, collide, or become entangled, collections of them appear as water or fire, plant or man. (fragment Diels-Kranz 68 A57, quoted from Graham 2010, p. 537).

To turn to contemporary physics, Feynman says at the beginning of the famous *Feynman lectures*:

> If, in some cataclysm, all of scientific knowledge were to be destroyed, and only one sentence passed on to the next generations of creatures, what statement would contain the most information in the fewest words? I believe it is the *atomic hypothesis* (or the atomic *fact*, or whatever you wish to call it) that *all things are made of atoms – little particles that move around in perpetual motion, attracting each other when they are a little distance apart, but repelling upon being squeezed into one another*. In that one sentence, you will see, there is an enormous amount of information about the world, if just a little imagination and thinking are applied.
> (Feynman et al. 1963, ch. 1-2)

What makes atomism attractive is evident from these quotations: on the one hand, it is a proposal for an ontology of nature that is most parsimonious and most general. On the other hand, it offers a clear and simple explanation of the realm of the objects that are accessible to us in perception. Any such object is composed of a finite number of discrete, pointlike particles. All the differences between these objects – at a time as well as in time – are accounted for in terms of the spatial configuration of these particles and its change. This view is implemented in classical mechanics. It conquered the whole of physics via classical statistical mechanics (e.g. heat as molecular motion), chemistry via the periodic table of elements, biology via molecular biology (e.g. molecular composition of the DNA), and finally neuroscience – neurons are composed of atoms, and neuroscience is applied physics (applied classical mechanics and electrodynamics, or, maybe, quantum mechanics in case quantum effects are proven to be operational in the brain).

What is relevant for the account of the perceptible macroscopic objects are only the relative positions of the point particles – that is to say, how far apart they are from each other, i.e. their distances – and the change of these distances. Let's call the particles "matter points". They can be considered as substances because they persist. However, in contrast to almost all the traditional philosophical accounts of substances, they are not Aristotelian substances in the sense of objects that have an inner form (eidos) – in other words, that are characterized by some intrinsic properties. There is nothing more to the matter points than the way in which they are spatially arranged and the change in their arrangement. But that is sufficient for their individuation: each matter point can be distinguished by – and hence individuated by – the distances it bears to the other matter points in a given configuration.

More precisely, if there is a configuration of $N$ matter points $i, j, k \ldots$, there are $1/2(N–1)$ distance relations. These relations are irreflexive, symmetric and connex (meaning that all matter points in a given configuration are related with one another). They satisfy the triangle inequality. For these relations to individuate the matter points, we have to stipulate that if matter point $i$ is not identical with matter point $j$, then the two sets that list all the distance relations in which these points stand with respect to all the other points in a configuration must differ in at least one such relation. We thereby exclude *entirely* symmetrical configurations among others. This is a structural individuation of the physical objects by relations in contrast to intrinsic essences. It has the great advantage that we do not have to endorse the numerical plurality of matter points as a primitive fact, which would imply that the matter points are bare particulars or bare substrata. Instead, they are individuated by the distance relations. To put it in a nutshell, matter, consisting in matter points, is what is individuated by its standing in distance relations to each other (by contrast to minds, angles, or abstract objects, which, if they exist, do not stand in spatial relations).

We can thus sum up the gist of atomism in these two axioms (for a detailed argument, see Esfeld and Deckert 2017, ch. 2.1, as well as Esfeld 2017 for a concise metaphysical argument):

(1) *There are distance relations that individuate simple objects, namely matter points.*
(2) *The matter points are permanent, with the distances between them changing.*

Let's call the ontology of nature defined by these two axioms the *primitive ontology*: matter points individuated by distances and their change are the ultimate referent of our physical theories, the bedrock of nature according to science so to speak.

However, the idea of matter being constituted by atoms in the sense of matter points is not sufficient to fulfil the promise of atomism, namely to explain everything in nature on the basis of the atomic hypothesis. That explanation is not carried out by the hypothesis of the atomic constitution of matter as such, but by showing how the change in the atomic composition of macroscopic objects accounts for their perceptible change. In other words, in order to fulfil the promise of atomism, one has in the first place to provide for laws of motion of the matter points and then to show how from these laws one also gets to an explanation of the motions of the macroscopic objects with which we are familiar. But the conceptual means provided by the primitive ontology – that is, the concepts of matter points, distances and their change admitted as primitive – are not sufficient to formulate a law of motion. Using only these conceptual means, one could not do much better than just listing the change that actually occurs, but not formulate a simple law that captures that change. The reason is that there is nothing about the distance relations in any given configuration of matter that provides information about the evolution of these relations.

To extract such information from the configuration of matter, we have to embed that configuration in a geometry and a dynamics: we have to conceive the configuration of matter as being embedded in a space with a fully-fledged metric (such as three-dimensional Euclidean space) – although in the ontology, there are only distance relations and their change, but not an absolute space or space-time. Furthermore, we have to attribute parameters to the configuration of matter that are introduced in terms of their functional role for the change in the distance relations. These can be parameters that are attributed to the matter points individually (such as mass, momentum, charge), to their entire configuration (such as total energy, or an initial wave function), or constants of nature (such as the gravitational constant). They can always remain the same (such as mass and charge) or vary as the distance relations among the matter points change (such as momentum, a wave function, etc.). In any case, conceiving the configuration of matter as being embedded in a geometrical space and as being endowed with parameters that are set up in terms of their function for the change in the distance relations then enables the formulation of a physical law. Let us call these parameters and the geometry, providing for physical laws, the *dynamical structure* of a physical theory. In fact, the geometry, the dynamical parameters and the laws come as a package: the precise functional definition of the dynamical parameters involves the law, and the law is formulated by using the dynamical parameters as well as the geometry. But there is no threatening circularity here: roughly speaking, all three are conjectured at once and then made precise together in order to achieve a theory that is simple and rich in informational content.

The claim then is that the primitive ontology remains constant – from Democritos to today's physics –, whereas the dynamical structure changes as we make more progress in formulating a theory that describes the evolution of the configuration of matter in a way that is ever more informative, while remaining as simple, general and informative as possible (for details, see Esfeld and Deckert 2017, ch. 2.2). In other words, there is something in a physical theory that serves as the – ultimate – referent of the theory, what there simply is in nature according to the theory. That something can be specified independently of the theory change in the history of science: atoms in the guise of matter points characterized by their relative positions and the change of these positions are that something. Furthermore, there is something in a physical theory that is introduced in terms of the role that it plays (i.e. its function) for the evolution of what there simply is according to the theory.

Thus, in classical mechanics, point particles characterized by their relative positions are what there simply is according to the theory – they have no further function in the theory apart from filling the place of the candidate for what simply exists in nature –, whereas the parameter of mass, for instance, is introduced in terms of what it does for the motion of the particles. As, for instance, Mach (1919, p. 241) points out when commenting on Newton's *Principia*, "The true definition of mass can be deduced only from the dynamical relations of bodies". That is to say, both inertial and gravitational mass are introduced through their dynamical role, namely as dynamical parameters that couple the motions of the particles to one another. In general, even if attributed to the particles taken individually, mass, charge, etc. express a dynamical relation between the particles instead of describing an intrinsic essence of the basic objects. As Hall (2009, § 5.2) puts it,

> the primary aim of physics – its first order business, as it were – is to account for *motions*, or more generally for change of spatial configurations of things over time. Put another way, there is one Fundamental Why-Question for physics: Why are things located where they are, when they are? In trying to answer this question, physics can of course introduce *new* physical magnitudes – and when it does, new why-questions will come with them.

This, again, alludes to the crucial distinction between primitive ontology and dynamical structure: the fundamental issue is the location of things and its change. The account of this fundamental issue requires the introduction of further parameters that allow us to formulate laws about how the change of location of things occurs.

Fields can with good reason be taken to belong also to the dynamical structure of physical theories instead of being part and parcel of the primitive ontology. In brief, (a) all the evidence for fields derives from evidence of particle motion. More importantly, (b) if one includes fields on a par with particles in the primitive ontology, the mathematical problem that there is no rigorous formulation of a physical theory of particle-field interactions, neither in classical electrodynamics nor in quantum field theory, becomes a philosophical problem how to conceive the interaction of these entities in the ontology. However, (c) if fields belong to the primitive ontology, their status is not clear: Are they properties of space-time points, albeit not geometrical ones? And why should only some space-time points have these causal properties (i.e. those where the field magnitudes are not zero)? Are they some sort of a bare substratum physical stuff? As Feynman puts it in his Nobel lecture,

> You see, if all charges contribute to making a single common field, and if that common field acts back on all the charges, then each charge must act back on itself. Well, that is where the mistake was, there was no field. It was just that when you shook one charge, another would shake later. There was a direct interaction between charges, albeit with a delay. … Now, this has the attractive feature that it solves both problems at once. First, I can say immediately, I don't let the electron act on itself, I just let this act on that, hence, no self-energy! Secondly, there is not an infinite number of degrees of freedom in the field. There is no field at all. (Feynman 1966, pp. 699-700; see Lazarovici 2018 for a detailed exposition of the arguments against a commitment to fields in the ontology of physics).

Also in what is known today as the standard model of particle physics in the framework of quantum field theory, the ontology of this physics can be set up in terms of a particle ontology only and the conceptual problems that this physics raises can thereby be answered (for a detailed account, see Esfeld and Deckert 2017, ch. 4).

In general, the benchmark for the dynamical structure of a physical theory is to simplify the representation of the change that takes place in the configuration of matter – by contrast to merely dressing a list of that change – without losing the information about the change that actually occurs. The common way to achieve this benchmark is to specify a dynamical structure such that, for any configuration of matter given as initial condition, the law fixes how the universe would evolve if that configuration were the actual one. The dynamical structure then goes beyond the actual configuration of matter: it fixes for any possible configuration of matter what the evolution of the universe would be like if that configuration were actual. It thereby supports counterfactual propositions.

Against this background, it is evident why determinism is a virtue of a physical theory: dynamical parameters figuring in laws that fix all the change, given an initial configuration of matter, are the simplest and most informative way to capture change. In other words, in the ideal case, the law is such that given an initial configuration of matter as input, the law yields a description of all the – past and future – change of the configuration as output. The question that remains in this case only is whether that description is empirically correct and whether it can be further simplified without losing informational content. It may turn out that, as a matter of fact, such a law cannot be achieved. It may also be that an indeterministic theory is simpler than a deterministic one and that the gain in simplicity outweighs the gain in informational content that a deterministic theory provides, such that, when seeking for the best balance between these two criteria, the indeterministic theory wins (see Werndl 2013 for a detailed elaboration on these issues). That notwithstanding, if there are dynamical parameters that designate only possibilities for how the configuration of matter may evolve, given an initial configuration, there always remains the question open whether one can do better, that is, find dynamical parameters that fix that change.

In any case, a fundamental physical theory is such that it defines a dynamical structure for the configuration of matter of the *whole* universe. For example, in Newtonian gravitation, the acceleration of any particle at any time depends, strictly speaking, on the positions and masses of *all* the other particles in the universe at that time. Even if action at a distance in Newtonian gravitation is replaced with local action in classical field theories, as soon as there are globally conserved quantities (such as total energy), the motion of any one object in the universe then is represented as being correlated with, in the end, the motion of any other object in the universe such that the quantity in question is globally conserved. In quantum physics, again, strictly speaking, due to entanglement, there is only one wave function for the configuration of matter as a whole at any given time (i.e. the universal wave function).

On the one hand, thus, the dynamical structure of a fundamental physical theory is defined for the universe as a whole. On the other hand, any such dynamical structure is *per se* useless for calculations. We cannot know initial conditions for the configuration of matter as a whole. Furthermore, the evolution of a given configuration of matter points that we can manipulate may be extremely sensitive to perturbations on its initial conditions. Hence, a slight error about the initial conditions may lead to a great error in predicting the evolution of the system. Already this fact makes clear that there is no conceptual link between deterministic laws and our ability to predict with certainty the evolution of a given system. Everything depends on the extent to which we can specify the initial conditions of a system and on how sensitive the evolution of the system is to slight variations of its initial conditions.

By way of consequence, setting out a primitive ontology and a dynamical structure is not sufficient to build up a physical theory. The dynamical structure has to be construed in such a way that it allows us to answer the following question: What evolution of a given system can we typically expect – that is, in the vast majority of situations – under ignorance of its exact initial conditions? For instance, when flipping a coin, it is impossible to predict the individual outcomes and thus to predict the exact sequence of heads and tails, although this sequence is completely determined by the laws of classical mechanics and the initial conditions. Nevertheless, it is possible to derive the proposition that in by far the most cases, the number of heads will be almost equal to the number of tails provided that the number of coin flips is large enough. There are situations in which we can predict individual outcomes, such as when throwing a stone on Earth, but these are the exception rather than the rule. The dynamical structure of a physical theory therefore has to be linked with a typicality or probability measure by means of which we can derive propositions about which evolution of particular systems we can expect to obtain in most cases under ignorance of the exact initial conditions. There hence is a clear reason why even a deterministic physical theory requires probabilities and a detailed procedure how to introduce them on the basis of – fundamental and universal – laws (for details, see Esfeld and Deckert, ch. 3.4).

As regards classical mechanics, notably Boltzmann has established how to derive such probabilistic statements from the deterministic laws via a typicality or probability measure (see Lazarovici and Reichert 2015 for a detailed account). Classical statistical mechanics then paved the way for developing atomism into precise scientific theories also in chemistry, biology and beyond. As regards quantum mechanics, it is by no means evident that the situation with respect to probabilities is different from the one in classical physics. It is a fact that situations like the classical coin toss are generic in quantum mechanics – that is, situations that are highly sensitive to slight variations of the initial conditions, and we cannot know these initial conditions with arbitrary precision. This fact is brought out by Heisenberg's uncertainty relations. Consequently, we can only make predictions about the statistical distributions of measurement outcomes by using Born's rule, but in general not predictions about individual measurement outcomes.

However, this fact does not imply that probabilities have another status in quantum mechanics than in classical mechanics. The question is what the law of motion for the evolution of the individual quantum systems is that underlies Born's rule for the calculation of measurement outcome statistics. Only if one includes what is introduced in the textbook presentations of quantum mechanics as the postulate of the collapse of the wave function upon measurement into the law does one obtain an indeterministic law in quantum mechanics. Doing so requires amending the Schrödinger equation with parameters that include the collapse of the wave function under certain circumstances. As things stand, these parameters have to be introduced by hand and compromise the simplicity of the law (see Ghirardi, Rimini and Weber 1986). Furthermore, they lead to predictions that deviate from the textbook ones in certain specific situations. In any case, it is an open issue whether such an indeterministic law is a fundamental or rather a phenomenological one – taking gravitation into account, for example, may turn this law into a deterministic one (see Penrose 2004, ch. 30). The only example of a candidate for an indeterministic law in a fundamental physical theory hence confirms the general statement made above, namely that in the case of an indeterministic law, it remains an open issue whether that law can be turned into a deterministic one by including further parameters.

Apart from the version of quantum mechanics that includes the postulate of the collapse of the wave function in the physical law, there are two other versions that both are deterministic. In brief, the version going back to Everett (1957) admits only the Schrödinger equation and, in consequence, no unique measurement outcomes. It is therefore known as many worlds quantum mechanics, because, in short, every possible outcome of a measurement becomes real in a branch of the universe (see Wallace 2012 for details). The version going back to Bohm (1952) adds to the (deterministic) Schrödinger equation a further (deterministic) law, known as the guiding equation, that describes, in brief, how the particles move in physical space as guided by the wave function. In the elaboration of this theory known as Bohmian mechanics, it is shown how Born's rule can be deduced from these laws by means of a typicality or probability measure that is linked with these laws in a way that matches the way in which the probability calculus of classical statistical mechanics is deduced from the deterministic laws of classical mechanics (see Dürr et al. 2013, ch. 2). The existence of Bohmian mechanics hence refutes any attempt to infer from Born's rule – or the Heisenberg uncertainty relations, or the randomness of individual measurement outcomes – the conclusion that probabilities have a more fundamental status in quantum mechanics than in classical mechanics. The question is what the law is that underlies Born's rule. The standard for assessing the proposals for that law is independent of the issue of determinism vs. indeterminism. The standard is what is the best solution to the quantum measurement problem (as illustrated, for instance, in Schrödinger's cat paradox). There are cogent arguments in favour of the Bohmian solution to this problem (see e.g. Esfeld 2014). The consequence then is that probabilities in quantum physics have the same status as probabilities in classical physics.

To sum this section up:

(1) There is a clear reason to seek for deterministic laws in the formulation of a physical theory, since these maximize informational content and usually also simplicity.

(2) There is a clear procedure available how to get from fundamental deterministic laws to predictions about statistical distributions of measurement outcomes both in classical and in quantum physics.

(3) Apparently random behaviour of investigated systems (including rules stating that randomness, such as the Heisenberg uncertainty relations) never justifies the conclusion to indeterminism. The issue is what the laws underlying this behaviour are. It is true that the determinism in classical mechanics would lose persuasiveness if there were not the clear cut paradigm examples of deterministic predictions in classical gravity (such as throwing a stone on Earth), and it is a fact that there are no such clear cut cases in quantum mechanics. But this is merely a heuristic matter. There is no conceptual link from deterministic laws to deterministic predictions, and, hence, no link from probabilistic predictions to probabilistic laws either.

## 2.    *Explanations in physics*

The *raison d'être* for laws in physics is that they explain the observed phenomena by subsuming them under a law – in whatever way one then spells out in philosophy of science how bringing phenomena under a law explains them (covering law, causal explanation, unification, just to name the most prominent accounts). This role of the laws raises the issue of their ontological status. In any case, as regards our knowledge, we cannot but make conjectures about what the laws are based on the data that become available to us. The standard for these conjectures is the extent to which they optimize both simplicity and informational content in accounting for the data. According to the stance known as Humeanism in today's metaphysics, this is all there is to the laws: they are nothing more than means of representation that seek to optimize simplicity and informational content. Super-Humeanism goes beyond standard Humeanism (see e.g. Lewis 1986, introduction) by putting the geometry and the dynamical parameters – that is, the dynamical structure – also on the side of the laws: the ontology is only the primitive ontology, such as matter points individuated by distance relations and the change in these relations. That change manifests certain patterns. Geometry, dynamical parameters and the laws linked with a probability measure are the package that enables us to achieve a representation of these patterns that is both as simple about the patterns and as informative about the change as possible (see Esfeld and Deckert 2017, ch. 2.3, for details).

(Super-)Humeanism is distinct from instrumentalism. It is a scientific realism: the claim is that what there is as far as the ontology of the natural world is concerned is exhausted by the primitive ontology. Dynamical parameters have a nomological role by figuring in the laws of nature. From that nomological role then derives their role in the predictions, as the laws are linked with a procedure to derive probabilities from them as sketched out in the previous section. The claim of Humeanism then is that the laws do not require additional ontological commitments. The claim of Super-Humeanism is that geometry, dynamical parameters and laws form a package that has only a representational purpose and that does hence not call for ontological commitments that reach beyond the primitive ontology. In short, the issue is what the ontology of the natural world is in a scientific realist framework.

Of course, physics explains the motions of bodies by using a geometry and dynamical parameters that appear in laws. However, the argument for an ontological commitment to the geometry and the dynamical parameters cannot simply be that they figure in our best physical theories. Reading the ontology off from the mathematical structure of physical theories would be begging the question of an argument for ontological commitments that go beyond what is minimally sufficient to account for the phenomena in a scientific realist vein, namely the commitment to a primitive ontology as given, for instance, by the two axioms of distance relations individuating matter points and the change in these relations. In a metaphysics based on science, the argument can only be that by subscribing to ontological commitments that go beyond that minimum, one achieves a gain in explanation.

(Super-)Humeanism can accommodate the scientific practice of explanations and its conceptualisation in terms of covering laws, causation or unification. There is no space in this paper to expand on this claim (see notably Loewer 2012 for details and the ensuing debate with Lange 2013, Miller 2015 and Marshall 2015). The core idea of the (Super-)Humean account is this one: the geometry and the dynamical structure of a physical theory explain the phenomena by bringing out the patterns or regularities in the motion of the particles; bringing out these patterns or regularities requires no ontological commitment beyond particles that move. On (Super-)Humeanism, first comes the particle motion, which as a contingent matter of fact exhibits certain patterns or regularities, then come the laws, including the geometry. Hence, the laws, the parameters figuring in them and the geometry are not some sort of an agent that forces the particles to move in a certain way. The laws do not constrain the particle motion. It is the particle motion that fixes the laws. Hence, if one asks why there are the patterns in the particle motion that there are in fact, (Super-)Humeanism cannot answer that question. The claim of (Super-)Humeanism is that there is no scientific answer to that question. Our scientific understanding of the world comes to an end once the salient patterns in the change of the elements of the primitive ontology are reached, such as, for instance, attractive particle motion.

The argument for this claim is the one illustrated in Molière's piece *Le malade imaginaire*: one does not explain why people fall asleep after the consumption of opium by endorsing a dormitive virtue of opium, because the dormitive virtue is *defined* in terms of its functional role to make people fall asleep after the consumption of opium. By the same token, one does not obtain a gain in explaining attractive particle motion by subscribing to an ontological commitment to gravitational mass as a property of the particles, because mass is *defined* in terms of its functional role of making objects attract one another as described by the law of gravitation. Of course, mass, charge and the like are fundamental and universal physical magnitudes, by contrast to the dormitive virtue of opium. But the point is that they are defined in terms of the functional role that they exert for the particle motion. Why do objects move as they do? Because they have properties whose function it is to make them move as they do. An ontological commitment to such properties does not yield a gain in explanation. The same holds for forces, fields, wave functions, an ontic structure of entanglement in quantum physics, laws conceived as primitive, etc. It also applies to geometry: it is no gain in explanation to trace the characteristic features of the distance relation back to the geometry of an absolute space, because that geometry is defined such that it allows for the conception of distances in that space.

It is true that by tracing the distance relations back to an absolute space, or the change in the distance relations back to properties of the particles that are dispositions for that very change, the characteristic features of the distance relations as well as those of the patterns in the change in them come out as necessary instead of contingent. However, merely shifting the status of something from contingent to necessary does not amount to a gain in explanation. Quite to the contrary, one only faces drawbacks that come with the commitment to a surplus structure in the ontology in the guise of an absolute space, fundamental dispositional properties of the particles, ontic dynamical structures of entanglement, etc.: differences with respect to absolute space that do not make a difference in the configuration of matter, questions such as how an object can influence the motion of other objects across space in virtue of properties that are intrinsic to it, how a wave function defined on configuration space can pilot the motion of matter in physical space, etc. (see Esfeld and Deckert 2017, ch. 2.3).

To sum this section up:

(1) The business of physics is to achieve on the basis of the available evidence a theory that is as simple and as informative as possible in accounting for that evidence and in predicting new evidence, with such a theory being characterized by the three features outlined in the previous section.

(2) Given the primitive ontology in terms of the notions of distances individuating matter points and the change of these distances, one can then define any further notion that one needs in one's theory of the natural world in terms of its functional role in the represenation of that change, without thereby subscribing to an additional ontological commitment. An ontology that is limited to a primitive ontology of matter points individuated by distance relations and the change in these relations is a scientific realism that is sufficient to accommodate scientific explanations.

(3) Subscribing to an ontological commitment that goes beyond what is minimally sufficient to account for the evidence (i.e. the primitive ontology) is not implied by the physics: one cannot read off the ontology from the mathematical structure of a physical theory. The issue can only be whether granting that structure an ontological status over and above the primitive ontology yields an explanatory gain. However, far from doing so, such an enriched ontology leads only to drawbacks stemming from a commitment to surplus structure.

## 3.    *Free will and the limits of physics*

Minimizing the ontological commitments of physics as outlined in the two preceding sections, while fully respecting scientific realism, not only prevents artificial problems from arising in the philosophy of nature, but also has repercussions for metaphysics in general. In particular, against the background set out here, one can establish the conclusion that there is no conflict between physical determinism and free will – although, at first glance, there obviously seems to be such a conflict.

Suppose that classical mechanics were the correct physical theory of the universe. Then, given an initial state of the particle motion throughout the whole universe (which includes the attribution of values of mass to the particles over and above initial positions and velocities) and the laws of classical mechanics, the entire evolution of the universe is fixed by the laws – that is, the entire *future* evolution from that state on as well as the entire *past* evolution leading to that state; that is why this can be an initial state at an arbitrary time. Of course, already in classical mechanics, as pointed out at the end of section 1, nobody within the universe could know its initial state at any time with enough precision to turn the determinism implemented in the laws into predictions.

If one considers physical determinism to be troublesome when it comes to human free will, a little reflection shows that the determinism implemented in the dynamical structure of classical mechanics is not the reason for the trouble. The reason is the very fact of there being universal physical laws. Suppose that a version of quantum mechanics that includes what is known as the collapse of the wave function in the fundamental dynamical law were the correct physical theory of the universe (such as the theory of Ghirardi, Rimini and Weber 1986 mentioned at the end of section 1) and that the collapse of the wave function is an irreducibly stochastic process. Nevertheless, the dynamical law then fixes objective probabilities for wave function collapse to occur such that, given an initial quantum state of the universe at an arbitrary time that includes an initial wave function of the universe, several possible future evolutions of the universe are fixed with objective probabilities attached to them. If the decisions of human beings concerning the motions of their bodies can influence neither the initial state of the universe nor the deterministic laws of classical mechanics (on the supposition that they are the correct laws of the universe), they cannot influence the objective probabilities implemented in a fundamental stochastic law and an initial wave function either (on the supposition that wave function collapse is stochastic and included in the fundamental dynamical law of the universe) (see Loewer 1996). Hence, supposing that there is a conflict between deterministic physical laws and free will, one could not draw any profit for free will if the laws were indeterministic. If there is such a conflict, it concerns the very fact of there being universal physical laws, be they deterministic or not.

The common formulation that, in the case of determinism, the laws plus the initial conditions fix the entire evolution of the objects to which they apply may suggest that the laws somehow bring about the evolution of these objects. However, if this were so, the laws would bring about the *past* evolution of the objects from an arbitrary initial state back as well as the *future* evolution of the objects from an arbitrary initial state on. But no one thinks that the fact that given an initial state and a deterministic dynamical law, the past evolution leading to that state is fixed implies that the law brings about the past evolution by retrocausation. Hence, the mere statement of determinism contains no reason to think that the law brings about the future evolution either. A better formulation of determinism that avoids any ontological connotation of the verb "fix" therefore is this one: the propositions stating the laws of nature and the propositions describing the state of the world at an arbitrary time $t$ (i.e. the propositions describing the initial conditions) entail the propositions describing the state of the world at any other time. Thus formulated, it is clear that determinism in science is – only – about entailment relations among propositions. The question then is, supposing that determinism is true, what it is in the world that makes these propositions true, that is, in virtue of what in the ontology these entailment relations among propositions hold.

On the conception of physical laws sketched out in the preceding section, there can be no clash between laws of nature and free will (in whatever way one may conceive free will). The reason is, in brief, that the motion comes first, including the motion of our bodies that is the expression of our intentions, then come the laws. In other words, what makes the propositions that state the laws true is the *entire* motion of the objects in the universe, that is, the change that actually occurs throughout the entire evolution of the configuration of matter of the universe. If the laws are "mere patterns in the phenomena", as Hall (2009, p. 1) puts it, they do not govern or constrain those phenomena, let alone bring them about. Hence, in this case, there is no clash between laws of nature and human free will possible, since the bodily movements that humans choose to make are part of the phenomena. The laws are there to achieve an account of the motions that actually occur that is both maximally simple and maximally informative. Consequently, the laws do not predetermine our actions, they only represent what happens in nature (see Beebee and Mele 2002).

Only if one loads the laws of physics with some sort of necessitation – such as by conceiving them as modal primitives, tracing them back to fundamental dispositions, powers or modal ontic structures instantiated by the physical objects – can a conflict with free will ensue (at least on an incompatibilist conception of free will); there then is something in the world that is independent of our decisions and that makes our decisions necessary. However, as far as the ontology of physics is concerned, there is no need to subscribe to any such commitment, and doing so leads only to drawbacks, as argued in the previous section.

Consider the famous consequence argument by means of which van Inwagen seeks to establish a conflict between free will and determinism:

> If determinism is true, then our acts are the consequences of the laws of nature and events in the remote past. But it is not up to us what went on before we were born, and neither is it up to us what the laws of nature are. Therefore the consequences of these things (including our pre-sent acts) are not up to us. (van Inwagen 1983, p. 16)

On the view defended in this paper, the statement "it is not up to us what went on before we were born" is ambiguous if it refers to the initial state of the universe and the statement "neither is it up to us what the laws of nature are" is, strictly speaking, not quite correct. The latter statement is not quite correct on any version of Humeanism about laws, for the just mentioned reason that what goes on in the universe comes first and then come the laws. However, it would be implausible to take this to imply that if a person had chosen to do otherwise, the laws of nature would have been different.

Here again the virtues of Super-Humeanism show up: it is not only the laws, but the entire dynamical structure of the correct physical theory of the universe that depend on the change in the universe as a whole. As argued in the first section, all the dynamical parameters that are introduced in terms of their functional role for the change in the primitive ontology – that is, the particle motion – are there to simplify, that is, to achieve a representation of the particle motion that is as simple and as informative as possible. Thus, they are not intrinsic to the particles or their configuration at any time. That is to say: the state of the universe at any given time, which enters as initial condition into the laws, contains elements that are not intrinsic to what there is at that time, but depend on the overall change in the universe. These are notably the initial values of parameters such as mass, fields, the universal wave function, etc. In order for these parameters to play their role to simplify the account of the motion that actually occurs in the universe, what role these parameters play and, notably, what their initial values are, depends on the change that actually occurs in the universe – that is, to stress again, the correct value of these parameters that enters into the state of the universe *at any given time* depends not only on what motion happens in the universe earlier than that time, but also on what happens *later* than that time. To put it in a nutshell, we do not know the initial wave function of the universe not only because of a principled limit on our knowledge, but also because what is the *initial* wave function of the universe is only fixed at the *end* of the universe so to speak (because it depends on what the particle motion during the evolution of the universe turns out to be like).

Consequently, if human beings chose to do otherwise, in the first place, slightly different initial values for the dynamical parameter at the initial state of the universe would have to be figured out in order to achieve a system that maximizes both simplicity and informational content about the change that actually occurs in the universe. For the sake of illustration, assume that quantum physics is the correct theory of the universe. Then, what would be slightly different if humans chose to do otherwise than they actually did were not the Schrödinger equation and the Bohmian guiding equation or the GRW collapse law in the first place, but the universal wave function, that is, the values that this wave function takes as initial condition. In that way, van Inwagen's consequence argument turns out to be invalid without the Humean being committed to saying that it is up to us what the laws of nature are. Instead, there is an ambiguity in the phrase "it is not up to us what went on before we were born". If that phrase is to include reference to an initial state of the universe before we were born, then that initial state, insofar as it enters into a law of nature, includes values of parameters that are not intrinsic to that state, but that depend on what happens later in the universe, including the particle movements that are expressions of human free will.

Hence, this paper is directed against a certain sort of a scientific worldview, namely one that implies a misconception of the enlightenment that comes with science: it is not that science teaches us that if there are deterministic laws in physics – or, for the sake of the argument, deterministic laws in genetics or evolutionary biology –, our decisions are necessitated by factors that are outside of our control. In general, this paper is about limits of science when it comes to the central metaphysical questions. In contrast to other attempts in that sense that argue for a limitation of the range of physical laws within the physical domain itself (see e.g. von Wachter 2015 according to whom physical laws, even when they are deterministic, indicate only tendencies for what happens in nature), the argument of this paper takes universal physical laws, also when they are deterministic, at face value as encompassing all the motions of bodies in the universe in a simple and general equation (or at least as striving for that ideal, as illustrated by the Newtonian law of gravitation). The argument then is that attributing a modal status to these laws is not justified by the physics, even if scientific realism is taken for granted. From that then follow certain limits of science, in particular that there is no clash between the scientific representation of the motions of bodies in terms of universal and deterministic laws and some such motions being the manifestation of human free will.

Once one has identified a primitive ontology of the natural world and thus settled for the concepts admitted as primitive that characterize that ontology, it is possible to define every further concept that enters into one's theory of the world in terms of the function for the primitive ontology. This applies not only to the parameters that appear in physical theories, but to any concept, including the ones describing the mind. It is at least since Lewis (1972) well known how to provide a scheme for the functional definition also of mental concepts in terms of, in the last resort, changes of the physical configuration of the body and its environment. Such functional definitions are undisputed in the natural sciences: it would be odd, for instance, to postulate a heat stuff to account for thermodynamical phenomena, since these can be defined functionally in terms of changes in molecular motion. By the same token, it would be odd to postulate an *élan vital* to account for organisms and their reproduction. Again, since the advent of molecular biology, the evolution of organisms and their reproduction can be accounted for in terms of molecular biology. There is no explanatory gap here.

However, when it comes to consciousness as well as rationality and the normativity and free will that are linked to rationality, one may maintain that there is an explanatory gap in the sense that functional definitions in terms of, in the last resort, changes in the configuration of matter do not capture what is characteristic of mental phenomena (see Levine 1983). Once one has understood the science, it is obvious how a functional definition of, for instance, water in terms of the effects on the interaction of $H_2O$ molecules captures and explains the phenomenal features of water and how a functional definition of organisms captures and explains their reproduction, including the link from genotypes to phenotypes. However, it is not obvious – at least not obvious in the sense of these paradigmatic examples – what the qualitative character of conscious experience, or the normativity that comes with rationality have to do with molecular motion in the brain.

The argument of this paper implies the following: in case the mental cannot be functionally defined on the basis of a primitive ontology of matter in motion, then an ontological commitment to the mental is called for over and above the ontological commitment to a primitive physical ontology. Moreover, such an ontological commitment then is as fundamental as the commitment to a primitive physical ontology, although the mental may only become manifest in certain systems in the universe and only at a certain period of time in the evolution of the cosmos. In general, whatever does not come in as being entailed by the primitive ontology via functional definitions is itself a further fundamental ingredient of the ontology (cf. e.g. Jackson 1994, or Chalmers 2012, although the argument of this paper is not committed to *a priori* entailment). This makes (again) evident the price that comes with any position whose ontological commitments go beyond a primitive physical ontology.

In the case of the additional parameters figuring in scientific theories, there is no reason to pay that price, as argued in the previous section. But the case of the mental is different. Positions that seek to avoid paying that price for instance by putting their stakes on emergence do not cut the ontological ice: if what emerges can be functionally defined on the basis of the ontology that is admitted as primitive, then there is no emergence in the sense of something that calls for new ontological commitments. If what emerges cannot be thus defined, then one is committed to more in the ontology than the ontology originally admitted as primitive. Consequently, there then are further primitives that hence have the same ontological status as the original primitives.

This is the core metaphysical debate, about the cosmos and about our place in it. Science can be understood on the basis of a primitive ontology that, even if the dynamics for that primitive ontology is deterministic, has no implications for what is right or wrong about these core metaphysical issues. Elaborating on the primitive ontology of science makes, however, clear the price that one has to pay for any further ontological commitments that then would have to come in as further primitives. The credibility of any such commitments hinges upon working them out into an overall metaphysical position that matches the paradigm of science in its clarity and precision as well as the concrete explanations that it provides.

Why determinism in physics has no implications for free will    15Hall, N. (2009): Humean reductionism about laws of nature. Unpublished manuscript, http://philpapers.org/rec/HALHRA

Jackson, F. (1994): Armchair metaphysics. In: Michael, M. – O'Leary Hawthorne, J. (eds.): *Philosophy in mind. The place of philosophy in the study of mind*. Dordrecht: Kluwer, 23–42.

Lange, M. (2013): Grounding, scientific explanation, and Humean laws. *Philosophical Studies* 164, 255–261.

Lazarovici, D. (2018): Against fields. *European Journal for Philosophy of Science* 8, 145–170.

Lazarovici, D. and Reichert, P. (2015): Typicality, irreversibility and the status of macroscopic laws. *Erkenntnis* 80, 689–716.

Levine, J. (1983): Materialism and qualia: the explanatory gap. *Pacific Philosophical Quarterly* 64, 354–361.

Lewis, D. (1972): Psychophysical and theoretical identifications. *Australasian Journal of Philosophy* 50, 249–258.

Lewis, D. (1986): *Philosophical papers. Volume 2*. Oxford: Oxford University Press.

Loewer, B. (1996): Freedom from physics: quantum mechanics and free will. *Philosophical Topics* 24, 91–112.

Loewer, B. (2012): Two accounts of laws and time. *Philosophical Studies* 160, 115–137.

Mach, E. (1919): *The science of mechanics: a critical and historical account of its development. Fourth edition. Translation by Thomas J. McCormack*. Chicago: Open Court.

Marshall, D. (2015): Humean laws and explanation. *Philosophical Studies* 172, 3145–3165.

Miller, E. (2015): Humean scientific explanation. *Philosophical Studies* 172, 1311–1332.

Penrose, R. (2004): *The road to reality: a complete guide to the laws of the universe*. London: Jonathan Cape.

van Inwagen, P. (1983): *An essay on free will*. Oxford: Oxford University Press.

von Wachter, D. (2015): Miracles are not violations of the laws of nature because the laws do not entail regularities. *European Journal for Philosophy of Religion* 7, 37–60.

Wallace, D. (2012): *The emergent multiverse. Quantum theory according to the Everett interpretation*. Oxford: Oxford University Press.

Werndl, C. (2013): On choosing between deterministic and indeterministic models: underdetermination and indirect evidence. *Synthese* 190, 2243–2265.